\documentclass[pra,onecolumn,superscriptaddress,nofootinbib,11pt,tightenlines]{revtex4}

\usepackage[a4paper,left=0.8in,right=0.8in,top=0.9in,bottom=0.9in]{geometry}
\usepackage{lmodern}
\usepackage{cmap}
\usepackage[utf8]{inputenc}
\usepackage[english]{babel}
\usepackage[T1]{fontenc}
\usepackage{microtype}
\usepackage{amsmath,amssymb,amsthm,mathrsfs,amsfonts}
\usepackage{mathtools}
\usepackage{bm}
\usepackage{booktabs,threeparttable,multirow}
\usepackage{graphicx}
\usepackage{float}
\usepackage[dvipsnames]{xcolor}
\usepackage{tikz}
\usetikzlibrary{arrows.meta,backgrounds,fit,positioning,calc}
\usepackage{natbib}
\usepackage{xurl}
\usepackage[
  bookmarks=false,
  colorlinks=true,
  urlcolor=BrickRed,
  linkcolor=BlueViolet,
  citecolor=PineGreen,
  pdfstartpage=1,
  pdfstartview={FitH}
]{hyperref}

\definecolor{agentblue}{HTML}{244A73}
\definecolor{agentteal}{HTML}{0F7C7B}
\definecolor{agentgold}{HTML}{C98B2E}
\definecolor{agentred}{HTML}{A63D40}
\definecolor{softblue}{HTML}{EAF1F7}
\definecolor{softteal}{HTML}{E8F4F3}
\definecolor{softgold}{HTML}{FBF2E3}
\definecolor{placeholderred}{HTML}{C62828}

\newcommand{\qcode}[3]{\ensuremath{\left[\!\left[#1,#2,#3\right]\!\right]}}

\renewcommand{\thesection}{\Roman{section}}
\renewcommand{\thesubsection}{\Alph{subsection}}

\makeatletter
\renewcommand{\p@subsection}{\thesection.}
\renewcommand{\p@subsubsection}{\thesection.\thesubsection.}
\makeatother

\begin{document}

\title{Multi-agent discovery of practical quantum LDPC codes}
\author{Dongheng Qian\textsuperscript{$*,\dagger$}}
\affiliation{State Key Laboratory of Surface Physics and Department of Physics, Fudan University, Shanghai 200433, China}
\affiliation{Shanghai Research Center for Quantum Sciences, Shanghai 201315, China}
\author{Tianyi Li\textsuperscript{$\ddagger,\dagger$}}
\affiliation{University of Wisconsin--Madison, Madison, Wisconsin 53706, USA}
\date[]{}

\begin{abstract}
Quantum low-density parity-check (qLDPC) codes can encode multiple logical qubits using sparse parity checks, yet searching for useful finite-length instances remains a challenging design problem because code performance must be optimized while satisfying practical constraints. Motivated by recent advances in artificial-intelligence agents for scientific discovery, we develop a multi-agent framework for discovering practical qLDPC codes. The framework combines specialist proposal and review, persistent scientific memory, long-horizon evolution of executable programs, and deterministic construction and evaluation within a closed-loop search. These programs instantiate coset-orbit balanced-product codes, providing a search space that includes bicycle and lifted-product constructions as well as non-normal subgroup actions. To incorporate practical constraints, we restrict the search to binary CSS codes with block length \(n\leq400\) and overall weight \(w\leq10\). Within this regime, the framework discovers codes with leading or competitive rate--distance performance in every weight class considered, with representative instances including \(\qcode{288}{16}{18}\) at \(w=7\), \(\qcode{288}{18}{18}\) at \(w=9\), and \(\qcode{234}{28}{18}\) at \(w=10\).  The search also uncovers structurally distinct, high-performing constructions, including a \(\qcode{336}{12}{\leq24}\) candidate and a \(\qcode{368}{18}{16}\) code, both of which are genuine balanced-product constructions with non-normal subgroup actions. When evaluated under code-capacity depolarizing noise using a common BP--OSD decoding protocol, the discovered codes also exhibit low logical failure rates. Together, these results provide hardware-relevant finite-length candidates for further experimental evaluation and show how structured agentic search can contribute to scientific discovery.
\end{abstract}

\maketitle
\begingroup
\renewcommand{\thefootnote}{\fnsymbol{footnote}}
\footnotetext[1]{Email: dhqian2022@gmail.com}
\footnotetext[2]{These authors contributed equally to this work.}
\footnotetext[3]{Email: tianyi.phys@gmail.com}
\endgroup

\section{Introduction}
Quantum error correction protects logical information by encoding it redundantly across physical qubits and repeatedly measuring parity checks to identify errors without directly measuring the logical state \citep{NielsenChuang2010}. Surface codes have become a leading route to fault-tolerant quantum computation because they combine comparatively high thresholds with geometrically local syndrome extraction in two dimensions \citep{Fowler2012Surface}. Recent experiments have demonstrated surface-code operation on superconducting and reconfigurable neutral-atom platforms \citep{Google2023Surface,Google2025BelowThreshold,Bluvstein2026Neutral}. The locality of surface code, however, comes with a vanishing encoding rate and consequently a large physical-qubit overhead. Quantum low-density parity-check (qLDPC) codes offer a complementary route: sparse parity checks are compatible with a finite encoding rate and substantially better asymptotic rate--distance trade-offs \citep{TillichZemor2014HGP,PanteleevKalachev2022Good,Dinur2022Good,LeverrierZemor2022Tanner,Bravyi2024Nature}. Realizing these advantages generally requires nonlocal interactions during syndrome extraction, but such connectivity is becoming achievable on experimental platforms, including reconfigurable neutral-atom arrays and superconducting processors with long-range couplers \citep{Bluvstein2024Logical,Bluvstein2026Neutral,Wang2026LowOverhead}. These advances create a path toward realizing qLDPC codes and exploiting their improved encoding efficiency in physical devices.

Although theoretical advances have established qLDPC families with progressively stronger asymptotic parameters, culminating in asymptotically good constructions with constant rate and linear distance \citep{TillichZemor2014HGP,HastingsHaahODonnell2021Fiber,BreuckmannEberhardt2021Balanced,PanteleevKalachev2022Good,Dinur2022Good,LeverrierZemor2022Tanner}, these asymptotic results provide limited guidance for choosing useful codes at finite block lengths accessible to current hardware. In practice, finite-length qLDPC codes have largely been developed within expert-designed construction families, including hypergraph-product, lifted-product and bivariate-bicycle codes \citep{TillichZemor2014HGP,PanteleevKalachev2021Finite,Bravyi2024Nature}. Promising instances are typically identified by enumerating or heuristically optimizing parameters within the selected family \citep{PanteleevKalachev2021Finite,Bravyi2024Nature,Liang2025Twisted,LinPryadko2024TwoBlock,Aydin2026Coset,Mian2026Multicycle,OkadaKasai2026FiniteField,Hong2026DesignRate,HirasakiLee2026Lifting}. The choice of construction space creates a central trade-off. A narrowly parameterized family may permit efficient enumeration but explores only a limited set of codes; a more general construction space admits a wider range of candidates, yet systematic optimization over that space remains difficult. Practical code design also involves several competing objectives: a finite-length code must balance encoding rate and distance with sparse checks, bounded qubit participation, decoder performance and feasible syndrome extraction. Candidate evaluation presents a further bottleneck: rank and sparsity are inexpensive to compute, whereas reliable distance and logical-error estimates require substantially greater computational effort. These considerations make it difficult to search broadly and efficiently for strong finite-length qLDPC codes under practical constraints.

Recent advances in artificial intelligence may offer a route to addressing these challenges. Reinforcement learning and Bayesian optimization have been used to discover small quantum codes and optimize selected qLDPC families \citep{Su2025RL,HeLiu2025LowWeightRL,Chengyu2026Bayes}, while language-model-guided evolution has searched executable generators for bivariate-bicycle and lifted-product codes \citep{CruzBenito2026Evolution,Liu2026SCE}. At the same time, AI-scientist systems have begun to use coordinated agents, external tools, persistent memory and iterative experimentation to organize longer and more structured scientific workflows \citep{Gottweis2026CoScientist,Ghareeb2026Robin,InternAgent2025,Feng2026InternAgent15}. 

Here we develop a multi-agent framework for finite-length qLDPC discovery that efficiently searches a general class of qLDPC constructions while optimizing code performance under multiple practical constraints.\footnote{During the finalization of this manuscript, we became aware of related work that also uses agents to discover QEC codes \citep{Yan2026OmniQEC}. While both studies use agents for QEC-code discovery, they differ in system architecture, optimization objective and search scope.} A council of researcher agents proposes and cross-reviews search directions, a curator agent converts their conclusions and accumulated evidence into persistent tasks and lessons, and long-horizon worker agents evolve executable code-family generators through repeated construction, probing and revision. Each worker’s output is passed to a deterministic pipeline that validates the construction, removes equivalent codes and computes a proxy score \(Q_{\mathrm{proxy}}=kd_{\mathrm{ub}}^2/n\), where \(d_{\mathrm{ub}}\) is a distance upper bound obtained using QDistEvol~\citep{Webster2026Distance}. Because reliable distance estimation is computationally expensive, the researcher and curator agents selectively promote promising candidates to higher-budget QDistEvol runs. This adaptive allocation concentrates computational effort on the candidates for which improved distance estimates are most informative, thereby increasing search efficiency.

We apply the framework to coset-orbit balanced products, a construction space that contains bicycle and lifted-product codes and also permits non-normal subgroup actions \citep{BreuckmannEberhardt2021Balanced}. Each executable family is represented at four levels: local terms, protograph shape, subgroup action and underlying group family. This representation directly generalizes the three-level structured concept evolution (SCE) developed for lifted-product codes \citep{Liu2026SCE}, retaining hierarchical mutation of executable family generators while introducing subgroup actions and thereby expanding the search space. To maintain broad coverage of this enlarged space, a MAP-Elites archive treats the underlying group family as an explicit niche dimension, together with subgroup type, code rate and overall weight. Parent selection combines the proxy score with niche coverage and candidate freshness, distributing exploration across different group families while continuing to refine strong candidates. 

Specifically, we search binary CSS codes with \(n\leq400\) and \(6\leq w\leq10\), where the overall weight \(w\) is defined as the larger of the maximum stabilizer weight and the maximum number of checks incident on any qubit. These bounds limit both the complexity of individual stabilizer measurements and the number of check interactions required by each qubit, making the resulting codes more amenable to implementation on realistic hardware. The search yields competitive finite-length parameters in every weight class considered. For selected candidates, we determine the code distance \(d\) rigorously using mixed-integer linear programming (MILP) and evaluate their final parameter quality using \(Q=kd^2/n\). Among codes with rigorously established distances in our comparison, the \(\qcode{288}{16}{18}\), \(\qcode{288}{18}{18}\) and \(\qcode{234}{28}{18}\) codes attain the largest \(Q\) values at \(w=7\), \(9\) and \(10\), respectively. The search also produces structurally distinct constructions, including a \(\qcode{336}{12}{\leq24}\) candidate over $\mathrm{PSL}(2,11)$ and a \(\qcode{368}{18}{16}\) code over $A_6\times\mathbb Z_2$, both arising from non-normal subgroup actions. Finally, under a common BP--OSD decoding protocol for code-capacity depolarizing noise, selected discoveries perform comparably to the bivariate-bicycle benchmarks, while the \(\qcode{400}{16}{\leq22}\) code performs better than both benchmarks across the sampled physical error rates.
\section{The multi-agent search framework}

\begin{figure}[t]
\centering
\includegraphics[width=0.90\textwidth]{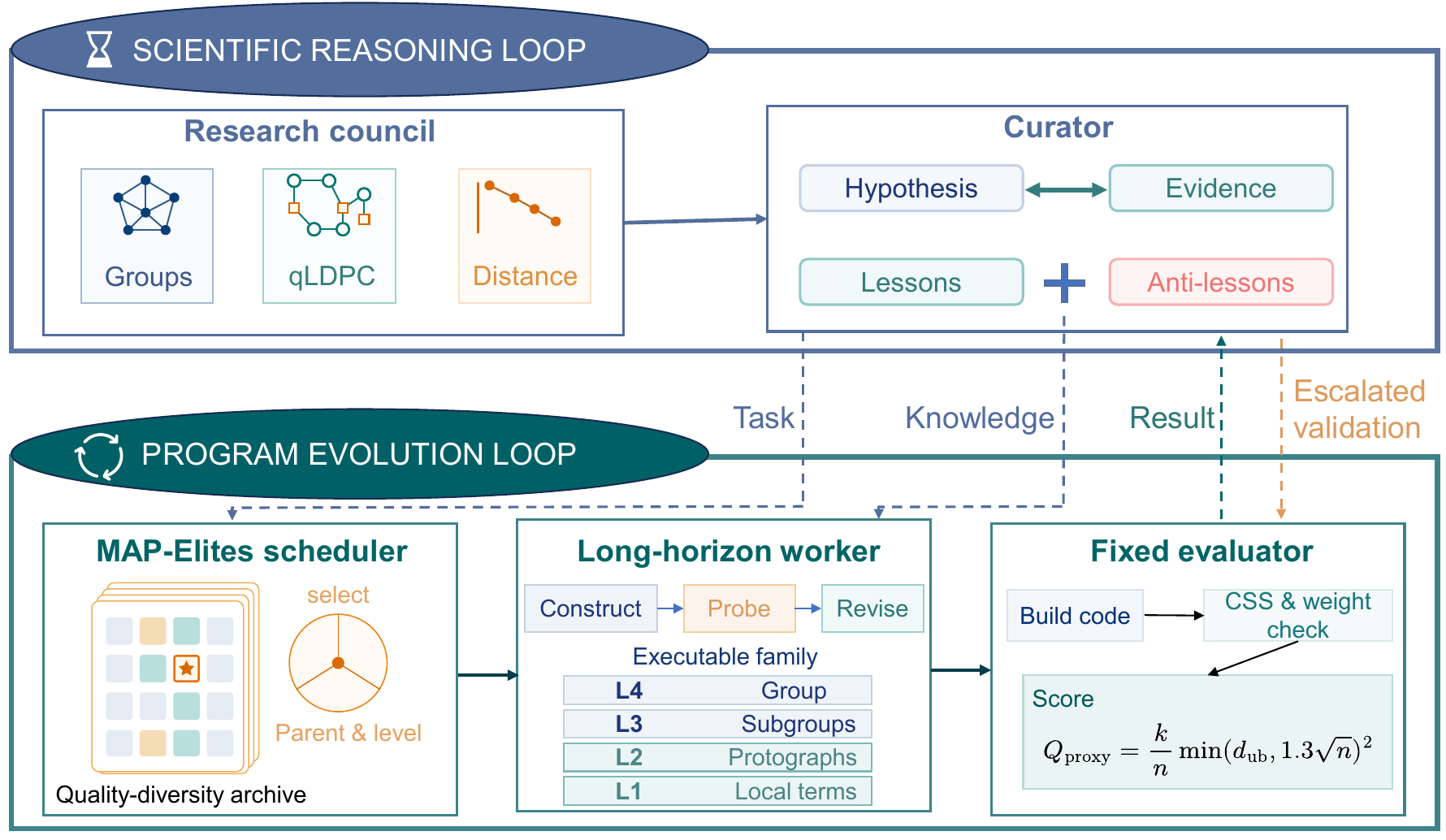}
\caption{\textbf{The multi-agent discovery framework.}
A scientific-reasoning loop (top) uses researcher proposals, curator synthesis and persistent memory to generate exploration and validation tasks. A program-evolution loop (bottom) mutates executable qLDPC-family generators within a MAP-Elites archive and scores candidates using a fixed evaluator. Evaluated candidates update both the archive and the evidence available to the scientific loop. L1--L4 denote mutations of local terms, protographs, subgroups and host groups, respectively.}
\label{fig:framework}
\end{figure}

The framework couples three agent roles to a fixed evaluator in a closed discovery loop, as summarized in Fig.~\ref{fig:framework}. A \emph{researcher council} formulates scientific directions, a \emph{curator} distills accumulated evidence into persistent knowledge and proposes new tasks, and a \emph{worker} executes those tasks by producing candidate programs. Together, the agents determine what should be explored, while the evaluator determines what constitutes an improvement. Inspired by IBM's evolutionary search for quantum codes~\cite{CruzBenito2026Evolution}, the objects evolved in this loop are executable programs that generate parameterized families of CSS codes, rather than individual parity-check matrices. This representation allows a promising modification to be explored simultaneously at several structural scales. Each program is indexed by a scaling parameter $t$, so different values of $t$ produce members of the same family at different block lengths.

The researcher council formulates hypotheses and proposes mathematical directions for the search. Its three members bring complementary expertise in finite groups and subgroup actions, qLDPC codes, and distance and sparse-graph structure. They first develop proposals independently and then examine one another's suggestions from their respective viewpoints, identifying changes that would make each proposal mathematically sound or more informative. The reviewed proposals are passed to the curator for synthesis.

The curator combines these theory-driven proposals with conclusions drawn from the search itself. It records a \emph{lesson} as a falsifiable relation between a design choice and an observed property, together with the candidates that support or contradict it. Positive lessons can progress from hypotheses to leaning and then confirmed statements as support accumulates, while recurring negative evidence is stored separately as an anti-lesson. Conversely, a counterexample can weaken an earlier lesson. By combining reviewed proposals, established lessons and unresolved contrasts, the curator formulates subsequent exploration and validation tasks.

After each worker submission, the fixed evaluator returns a quantitative outcome. It instantiates the submitted program at admissible values of $t$ in the block-length window $40\leq n\leq400$. A parameter point is retained only if $H_X H_Z^{\mathsf T}=0$, the number of encoded qubits $k=n-\operatorname{rank}(H_X)-\operatorname{rank}(H_Z)$ is positive and the LDPC weight constraint is satisfied. We define the overall weight as
\begin{equation}\label{eq:weight}
w=\max\Bigl\{\max_i\mathrm{wt}\bigl(H_{X,i}\bigr),\ 
                 \max_i\mathrm{wt}\bigl(H_{Z,i}\bigr),\ 
                 \max_q\bigl[\deg_X(q)+\deg_Z(q)\bigr]\Bigr\},
\end{equation}
where $H_{X,i}$ and $H_{Z,i}$ are check rows and $\deg_X(q)$ and $\deg_Z(q)$ count the $X$- and $Z$-type checks incident on physical qubit $q$. The search requires $w\leq10$. This definition constrains both stabilizer weight and the total number of check interactions involving a qubit.\footnote{This convention is more restrictive than definitions that characterize sparsity by the maximum check weight, or by the qubit degree within each Pauli sector separately. For example, the \qcode{300}{60}{14} code of Ref.~\citep{Bhardwaj2026Mitten} has check weight 9 and per-sector qubit degree at most 6, but each qubit participates in up to six $X$-type and six $Z$-type checks, so its overall weight under Eq.~\eqref{eq:weight} is 12.} The evaluator selects up to three representative values of $t$, prioritizing both large numbers of encoded qubits $k$ and high encoding rates $k/n$. Before scoring, it canonically labels the three-colored Tanner graph, whose vertex types distinguish physical qubits, $X$ checks and $Z$ checks, and discards codes isomorphic to an archived representative~\cite{junttila2007engineering}.

Each new representative is assigned the proxy score
\begin{equation}\label{eq:proxy-score}
Q_{\mathrm{proxy}}=\frac{k\widetilde d^{2}}{n},
\qquad
\widetilde d=\min\!\left(d_{\mathrm{ub}},1.3\sqrt n\right).
\end{equation}
Here $d_{\mathrm{ub}}$ is obtained by QDistEvol from an explicit logical operator and is therefore an upper bound on the true distance~\cite{Webster2026Distance}. Following IBM's evolutionary search~\cite{CruzBenito2026Evolution}, the cap in Eq.~\eqref{eq:proxy-score} prevents an unusually loose upper bound from dominating selection. The score assigned to a family is the largest credible value of $Q_{\mathrm{proxy}}$ among its fully evaluated representatives. Furthermore, because distance estimation dominates the cost of evaluation, it is allocated in stages. Each representative first receives a $10^4$-iteration QDistEvol search. If the resulting upper bound exceeds $1.3\sqrt n$, the evaluator automatically repeats the calculation at $10^5$ iterations before applying the credibility cap. Beyond this automatic escalation, we introduce a $10^6$-iteration tier that can be commissioned only through an explicit validation task issued by the curator, either directly or following a recommendation from the researcher council. Such a task is reserved for cases in which stronger evidence could resolve a lesson, distinguish competing explanations or alter a search decision. By directing the most expensive distance calculations towards candidates whose uncertainty matters to an active hypothesis or selection decision, this agent-guided allocation is expected to use the verification budget more efficiently than uniform escalation.

All of the design choices above are independent of the particular code construction. In the present search, we apply them to coset-orbit balanced-product CSS codes~\cite{BreuckmannEberhardt2021Balanced}. We choose this construction for two reasons. First, it admits a transparent hierarchical specification. A family is determined by four components: a finite host group $G(t)$; the assignment of subgroups to the protograph row and column types; the shapes of two protograph matrices $A(t)$ and $B(t)$ and their scaling with $t$; and the local terms in their entries. Each local term is an $\mathbb F_2$ sum of double-coset orbits $K_i g K_j$, where $K_i$ and $K_j$ are the subgroups assigned to the corresponding row and column types. A fixed assembler converts these four components into the binary check matrices $H_X(t)$ and $H_Z(t)$. These components directly define four mutation levels. L1 changes the local terms in $A$ and $B$; L2 changes the protograph shapes and their scaling with $t$; L3 changes the subgroup assignment; and L4 changes the executable construction of $G(t)$. This hierarchy extends the three-level representation used in lifted-product search~\cite{Liu2026SCE} by making the subgroup action an independent structural variable. Researcher and curator tasks specify the intended level, whereas automatic refinement tasks obtain one from the scheduler.

Second, balanced products provide a broad construction space in which strong qLDPC codes are expected to occur. On the one hand, a common trivial subgroup gives a lifted-product code over $G$~\cite{panteleev2022almost}, with generalized and bivariate bicycle codes arising as more specialized abelian cases~\cite{kovalev2013quantum,Bravyi2024Nature}. This region already contains asymptotically good qLDPC families~\cite{PanteleevKalachev2022Good,LeverrierZemor2022Tanner}. On the other hand, non-normal subgroup actions extend the construction beyond lifted products. A common normal subgroup reduces to a lifted product over $G/K$, so a genuinely beyond-LP construction requires a non-abelian host group and a non-normal subgroup. Allowing distinct subgroups for different row and column types broadens this less-explored region further.

Evaluated families are stored in a MAP-Elites archive, which partitions the population into niches and retains the best member of each niche~\cite{mouret2015illuminating}. The niche coordinates are the host-group family, subgroup structure, code-rate bin and overall weight. Scientific tasks proposed by the curator or researchers take priority. When no such task is ready, the scheduler chooses a parent and mutation level by balancing family score, coverage of underexplored group and rate regions, and the freshness of the lineage. Structural mutations are emphasized earlier in the campaign and local refinement increasingly later. Together, niche preservation and score-informed scheduling balance exploration and exploitation: separate niches maintain coverage across distinct groups, while selection within those niches continually refines the strongest candidates.

After evaluation, the child updates the archive and its contrast with the parent becomes new evidence for the curator and researcher council. Repeating this cycle couples short-range program evolution to longer-range revision of scientific hypotheses. This is the principal methodological advance over AlphaEvolve-style systems, in which executable-program mutations are guided primarily by fitness feedback from previously evaluated candidates~\cite{novikov2025alphaevolve,li2026madevolve}. Our framework retains this effective form of program evolution, including its recent use in quantum-code search~\cite{CruzBenito2026Evolution,Liu2026SCE}, but adds an explicit scientific-reasoning cycle: hypotheses are formulated across code families, tested against accumulated evidence and converted into new experiments.

Finally, selected output matrices are examined in a separate post-search MILP calculation. A distance is reported as exact only when a logical witness gives the matching upper bound and the optimization has excluded every smaller nontrivial logical operator; otherwise, it remains an upper bound~\cite{gurobi}.

\section{Search results}

\subsection{Leading finite-length codes across weight classes}

\begin{figure}[H]
  \centering
  \includegraphics[width=\textwidth]{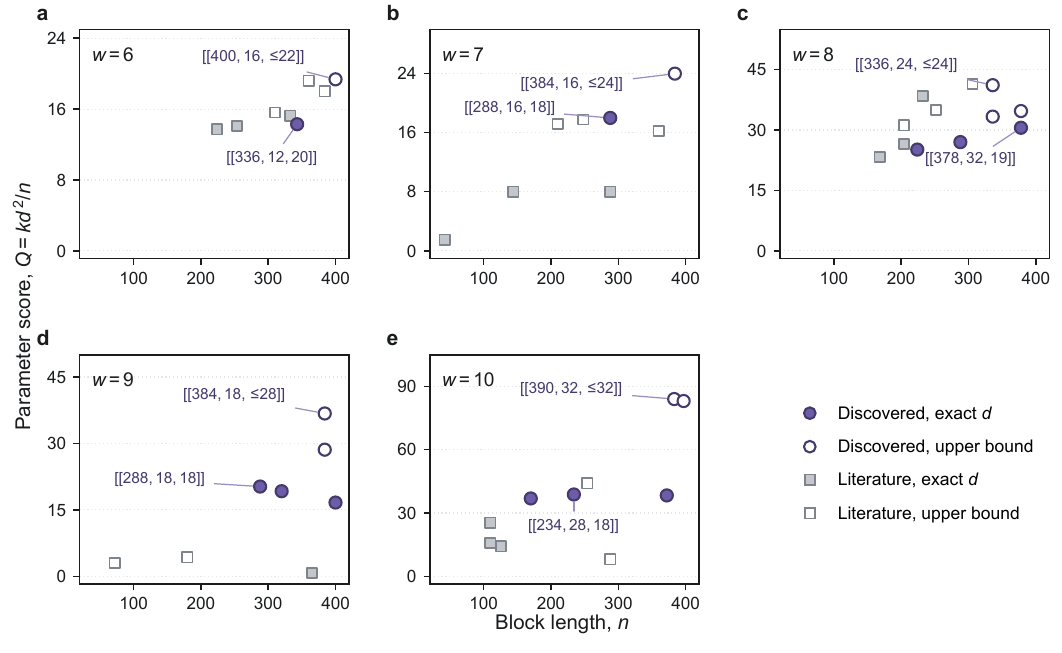}
  \caption{\textbf{Finite-length performance under overall-weight constraints.} Panels (a)--(e) compare the parameter score $Q=kd^2/n$ against block length $n$ at fixed overall weights $w=6$--$10$. Purple circles are codes discovered in this work and gray squares are literature codes; filled markers denote exact distances and open markers denote QDistEvol distance upper bounds, for which the plotted $Q$ is an upper endpoint. Selected discoveries from Table~\ref{tab:codes} are labeled with their $\qcode{n}{k}{d}$ parameters, including the exact-distance leaders \qcode{288}{16}{18}, \qcode{288}{18}{18} and \qcode{234}{28}{18} at $w=7$, $9$ and $10$.}
  \label{fig:parameter-landscape}
\end{figure}

The search produced competitive codes throughout the full range of sparsity constraints. Figure~\ref{fig:parameter-landscape} organizes the finite-length comparison into five panels, one for each overall-weight class. Because relaxing the weight constraint greatly changes the feasible construction space, $Q$ is compared at fixed $w$. Within each panel, rigorous distances and QDistEvol upper bounds are indicated by distinct markers.

The exact-distance results establish a strong finite-length frontier. At $w=6$, the \qcode{336}{12}{20} code attains $Q=14.29$, within 7\% of the leading exact code in the comparison, the \qcode{340}{16}{18} generalized-bicycle code with $Q=15.25$ \citep{Liang2025Twisted}. At $w=7$, the \qcode{288}{16}{18} code reaches $Q=18.00$. To our knowledge, this is the strongest known code in this weight class under $Q$; its rigorous score also exceeds the highest value derived from the distance upper bound, $Q\leq17.78$ \citep{Aydin2026Coset}. At $w=8$, the Heisenberg-group \qcode{378}{32}{19} code gives $Q=30.56$, while \qcode{288}{24}{18} and \qcode{224}{22}{16} codes obtained from distinct constructions give $Q=27.00$ and $25.14$. These values are competitive with the recently reported strongest code in this class \citep{OkadaKasai2026PairPartition}. The \qcode{288}{18}{18} code gives $Q=20.25$ at $w=9$ and also exceeds the previously known best result \citep{Jacob2025Tricycle}. At $w=10$, the \qcode{234}{28}{18} and \qcode{372}{44}{18} codes reach $Q=38.77$ and $38.32$, respectively, both surpassing the strongest previous exact code in this class \citep{OkadaKasai2026FiniteField}. Together, our certified discoveries provide best-known or competitive codes across every weight class considered.

For candidates whose putative distance is high enough that exact MILP certification becomes impractical, we performed $10^6$-iteration QDistEvol runs and report the smallest logical-operator weights found as distance upper bounds. At $w=6$, the \qcode{400}{16}{\leq22} code reaches $Q\leq19.36$. This endpoint exceeds $Q\leq19.20$ for the best-known weight-six bivariate-bicycle code, \qcode{360}{12}{\leq24} \citep{Liang2025Twisted}. The \qcode{384}{16}{\leq24} code raises the weight-seven endpoint to $Q\leq24.00$, while the \qcode{336}{24}{\leq24} code at $w=8$ reaches $Q\leq41.14$, close to the benchmark value $Q\leq41.41$ of \qcode{306}{22}{\leq24} \citep{Aydin2026Coset}. At $w=9$ and 10, the leading candidates are \qcode{384}{18}{\leq28} with $Q\leq36.75$ and \qcode{390}{32}{\leq32} with $Q\leq84.02$. Taken together, these upper-bound results indicate progress across all weight classes. They identify candidates that would substantially improve the known finite-length frontier, while further rigorous certification remains necessary.

Table~\ref{tab:codes} gives the construction data for the 20 discovered codes highlighted in Fig.~\ref{fig:parameter-landscape}. The leading codes span generalized-bicycle constructions, abelian lifted products, non-abelian metacyclic and dicyclic hosts, finite Heisenberg and modular groups, and a Frobenius construction. They vary not only the group family and subgroup action, including free and normal actions, but also the base-matrix shape, from $1\times1$ constructions to $3\times4$, $4\times4$ and $5\times5$ designs. The results therefore arise from mutations at every level of the executable representation, rather than from a large enumeration of local supports within a single established code family.

\begin{table}[H]
\centering
\begin{threeparttable}
\caption{\textbf{Finite-length codes selected from the search.} These 20 discoveries are the colored points in Fig.~\ref{fig:parameter-landscape}. Group family denotes the instantiated host $G$; $\rtimes$ denotes a semidirect product, with the subscript specifying the action when shown, $\mathrm{Dic}_t$ is dicyclic, $\mathrm{He}(\mathbb Z_3)$ is the finite Heisenberg group, $M_{27}$ is the modular group of order 27, and $\mathrm{SL}(2,3)$ and $\mathrm{GL}(2,3)$ are the special and general linear groups of degree two over $\mathbb F_3$. In $\mathrm{SL}(2,3)\rtimes_{\phi}\mathbb Z_{16}$, $\phi$ is the order-four automorphism induced by conjugation with $\left(\begin{smallmatrix}1&1\\2&1\end{smallmatrix}\right)\in\mathrm{GL}(2,3)$. Action gives the orbit stabilizer $K$, with ``free'' denoting $K=\{e\}$ and ``normal'' or ``non-normal'' specifying its relation to $G$. Shape gives the row-by-column dimensions of $A$ and $B$ as $A/B$. Numeric distances are rigorous; entries containing $\leq$ are QDistEvol upper bounds, with the corresponding $Q$ values reported as upper-bound endpoints.}
\label{tab:codes}
\scriptsize
\setlength{\tabcolsep}{3.2pt}
\begin{tabular}{@{}c l r p{0.25\textwidth} p{0.21\textwidth} l@{}}
\toprule
$w$ & Code & $Q$ & Group family & Action & Shape \\
\midrule
6 & \qcode{336}{12}{20} & 14.29 & $\mathbb Z_{84}\rtimes_{29}\mathbb Z_4$ & normal $|K|=2$ & $1\times1/1\times1$ \\
6 & \qcode{400}{16}{\leq22} & $\leq19.36$ & $\mathrm{Dic}_5\times\mathrm{Dic}_5$ & normal $|K|=2$ & $1\times1/1\times1$ \\
\addlinespace
7 & \qcode{288}{16}{18} & 18.00 & $\mathbb Z_{12}\times\mathbb Z_{48}$ & normal $|K|=4$ & $1\times1/1\times1$ \\
7 & \qcode{384}{16}{\leq24} & $\leq24.00$ & $\mathbb Z_{12}\times\mathbb Z_{48}$ & normal $|K|=3$ & $1\times1/1\times1$ \\
\addlinespace
8 & \qcode{378}{32}{19} & 30.56 & $\mathrm{He}(\mathbb Z_3)\times\mathbb Z_7$ & free & $1\times1/1\times1$ \\
8 & \qcode{288}{24}{18} & 27.00 & $\mathbb Z_{12}\times\mathbb Z_{48}$ & normal $|K|=4$ & $1\times1/1\times1$ \\
8 & \qcode{224}{22}{16} & 25.14 & $\mathbb Z_{84}\rtimes_{29}\mathbb Z_4$ & normal $|K|=3$ & $1\times1/1\times1$ \\
8 & \qcode{336}{24}{\leq24} & $\leq41.14$ & $\mathbb Z_4\times\mathbb Z_{42}$ & free & $1\times1/1\times1$ \\
8 & \qcode{378}{18}{\leq27} & $\leq34.71$ & $M_{27}\times\mathbb Z_7$ & free & $1\times1/1\times1$ \\
8 & \qcode{336}{28}{\leq20} & $\leq33.33$ & $\mathbb Z_2^2\times\mathbb Z_{42}$ & free & $1\times1/1\times1$ \\
\addlinespace
9 & \qcode{288}{18}{18} & 20.25 & $\mathrm{SL}(2,3)\rtimes\mathbb Z_{12}$ & normal $|K|=2$ & $1\times1/1\times1$ \\
9 & \qcode{320}{24}{16} & 19.20 & $\mathbb Z_{41}\rtimes\mathbb Z_8$ & normal $K=\mathbb Z_{41}$ & $4\times4/5\times5$ \\
9 & \qcode{400}{26}{16} & 16.64 & $\mathrm{Dic}_4$ & free & $3\times4/3\times4$ \\
9 & \qcode{384}{18}{\leq28} & $\leq36.75$ & $\mathrm{SL}(2,3)\rtimes_{\phi}\mathbb Z_{16}$ & normal $|K|=2$ & $1\times1/1\times1$ \\
9 & \qcode{384}{14}{\leq28} & $\leq28.58$ & $\mathrm{GL}(2,3)\times\mathbb Z_4$ & free & $1\times1/1\times1$ \\
\addlinespace
10 & \qcode{234}{28}{18} & 38.77 & $\mathbb Z_{13}\times\mathbb Z_9$ & free & $1\times1/1\times1$ \\
10 & \qcode{372}{44}{18} & 38.32 & $\mathbb Z_{31}\times\mathbb Z_6$ & free & $1\times1/1\times1$ \\
10 & \qcode{170}{32}{14} & 36.89 & $\mathbb Z_{85}$ & free & $1\times1/1\times1$ \\
10 & \qcode{390}{32}{\leq32} & $\leq84.02$ & $\mathbb Z_{195}$ & free & $1\times1/1\times1$ \\
10 & \qcode{390}{36}{\leq30} & $\leq83.08$ & $\mathbb Z_{39}\times\mathbb Z_5$ & free & $1\times1/1\times1$ \\
\bottomrule
\end{tabular}
\end{threeparttable}
\end{table}

\subsection{Exploration of nontrivial group constructions}

The preceding parameter comparison emphasizes exploitation, showing that the framework can locate high-quality codes under fixed sparsity constraints. We next examine the search from the complementary perspective of exploration, focusing on the diversity of nontrivial construction mechanisms among the discovered codes. While the codes in Table~\ref{tab:codes} all have free or normal $K$ and therefore reduce to lifted products over $G$ or $G/K$, the four-level representation broadens exploration within the lifted-product class. Independent mutations of the stabilizer subgroup and host group can make unusual lifted-product groups accessible through non-obvious host--subgroup descriptions. For example, in the \qcode{400}{16}{\leq22} code, the diagonal central stabilizer identifies the two central involutions $z_1$ and $z_2$ of $\mathrm{Dic}_5\times\mathrm{Dic}_5$. The resulting lifted product is defined over the order-200 central product
\begin{equation}
(\mathrm{Dic}_5\times\mathrm{Dic}_5)/\langle(z_1,z_2)\rangle.
\end{equation}
For the \qcode{384}{18}{\leq28} code, quotienting the host $\mathrm{SL}(2,3)\rtimes_{\phi}\mathbb Z_{16}$ by $K=\langle(-I,0)\rangle$ gives
\begin{equation}
(\mathrm{SL}(2,3)\rtimes_{\phi}\mathbb Z_{16})/K
\cong A_4\rtimes_{\bar\phi}\mathbb Z_{16},
\end{equation}
where $A_4$ is the alternating group on four letters and $\bar\phi$ is the induced order-four automorphism. These two codes show how the four-level language reaches uncommon non-abelian lifted products rather than merely scanning local supports over standard cyclic or abelian hosts.

Table~\ref{tab:structural-codes} presents two further lifted-product examples based on distinctive groups. The exact \qcode{256}{18}{16} code, with $Q=18.00$, is constructed over the wreath product $\mathbb Z_8\wr\mathbb Z_2$, in which $\mathbb Z_2$ exchanges two copies of $\mathbb Z_8$. The exact \qcode{384}{32}{16} code, with $Q=21.33$, is generated from the twisted-dihedral host $\mathrm D_6\rtimes_{\theta}\mathbb Z_8$ with the normal stabilizer $K=\langle(r^3,0)\rangle$. Here $\mathrm D_6=\langle r,s\mid r^6=s^2=e,\,srs=r^{-1}\rangle$ has order 12, and $\theta$ fixes $r$ while sending $s$ to $r^3s$. Quotienting by $K$ removes the twist because $r^3=e$ and hence $\theta(s)=s$, giving
\begin{equation}
\frac{\mathrm D_6\rtimes_{\theta}\mathbb Z_8}{K}
\cong \left(\mathrm D_6/\langle r^3\rangle\right)\times\mathbb Z_8
\cong S_3\times\mathbb Z_8.
\end{equation}
Together, these examples illustrate how the search explores nontrivial group structure even when the resulting codes remain lifted products.

Moreover, the framework discovered genuine balanced-product constructions with useful finite-length parameters.\footnote{Here ``genuine'' refers to the explicit $(G,K)$ presentation found by the search and does not rule out an alternative lifted-product representation of the same finite matrix.} For non-normal $K$, the coset space $G/K$ is not a quotient group, and the construction cannot be reduced through $G/K$ to an ordinary lifted product. The remaining five codes in Table~\ref{tab:structural-codes} realize four such mechanisms. The alternating-group construction uses $A_6\times\mathbb Z_2$, where $A_6$ is the alternating group on six letters, and yields the exact \qcode{368}{18}{16} code with a non-normal order-two stabilizer. Constructions over the projective special linear groups $\mathrm{PSL}(2,q)$ give the exact \qcode{248}{12}{18} code at $q=13$ and the \qcode{336}{12}{\leq24} candidate at $q=11$. A Frobenius construction over $\mathbb Z_{31}\rtimes\mathbb Z_6$ produces \qcode{396}{8}{\leq32}. Finally, a swap-semidirect construction over $(\mathbb Z_{17}^2)\rtimes_{\mathrm{swap}}\mathbb Z_2$ produces \qcode{306}{8}{\leq25} and admits the closed double-coset count, for which the number of K-double cosets has the closed form $t(t+1)/2$. These constructions span several distinct forms of non-normal group action, demonstrating that genuine balanced products are a recurring outcome of the exploration rather than an isolated exception.

Given the diversity of these genuine balanced-product examples, we conjecture that lifted products are intrinsically favored for maximizing $Q$ in the finite regime considered here, rather than appearing dominant because the framework failed to explore genuine balanced products. Under the simultaneous constraints $n\leq400$ and $w\leq10$, the additional orbit geometry of genuine balanced products has not yet translated into parameter scores comparable to the leading lifted products. Their advantages may emerge at larger block lengths or under less restrictive weight budgets.

\begin{table}[t]
\centering
\begin{threeparttable}
\caption{\textbf{Structurally distinct codes found by the framework.} These seven codes supplement the parameter-frontier selection in Table~\ref{tab:codes}, using the same column definitions. $A_6$ is the alternating group on six letters, $\mathrm{PSL}(2,q)$ is the projective special linear group over $\mathbb F_q$, $\wr$ denotes a wreath product, and $\mathrm D_6\rtimes_{\theta}\mathbb Z_8$ is the twisted-dihedral host defined in the text. Numeric distances are rigorous; entries containing $\leq$ are QDistEvol upper bounds, with the corresponding $Q$ values reported as upper-bound endpoints.}
\label{tab:structural-codes}
\scriptsize
\setlength{\tabcolsep}{3.2pt}
\begin{tabular}{@{}c l r p{0.25\textwidth} p{0.21\textwidth} l@{}}
\toprule
$w$ & Code & $Q$ & Group family & Action & Shape \\
\midrule
8 & \qcode{256}{18}{16} & 18.00 & $\mathbb Z_8\wr\mathbb Z_2$ & free & $1\times1/1\times1$ \\
8 & \qcode{384}{32}{16} & 21.33 & $\mathrm D_6\rtimes_{\theta}\mathbb Z_8$ & normal $|K|=2$ & $2\times2/2\times2$ \\
\addlinespace
9 & \qcode{368}{18}{16} & 12.52 & $A_6\times\mathbb Z_2$ & non-normal $|K|=2$ & $1\times1/1\times1$ \\
9 & \qcode{336}{12}{\leq24} & $\leq20.57$ & $\mathrm{PSL}(2,11)$ & non-normal $|K|=2$ & $1\times1/1\times1$ \\
\addlinespace
10 & \qcode{248}{12}{18} & 15.68 & $\mathrm{PSL}(2,13)$ & non-normal $|K|=3$ & $1\times1/1\times1$ \\
10 & \qcode{396}{8}{\leq32} & $\leq20.69$ & $\mathbb Z_{31}\rtimes\mathbb Z_6$ & non-normal $|K|=3$ & $3\times3/3\times3$ \\
10 & \qcode{306}{8}{\leq25} & $\leq16.34$ & $(\mathbb Z_{17}^2)\rtimes_{\mathrm{swap}}\mathbb Z_2$ & non-normal $|K|=2$ & $1\times1/1\times1$ \\
\bottomrule
\end{tabular}
\end{threeparttable}
\end{table}

\subsection{Code-capacity performance}

\begin{figure}[H]
  \centering
  \includegraphics[width=0.95\textwidth]{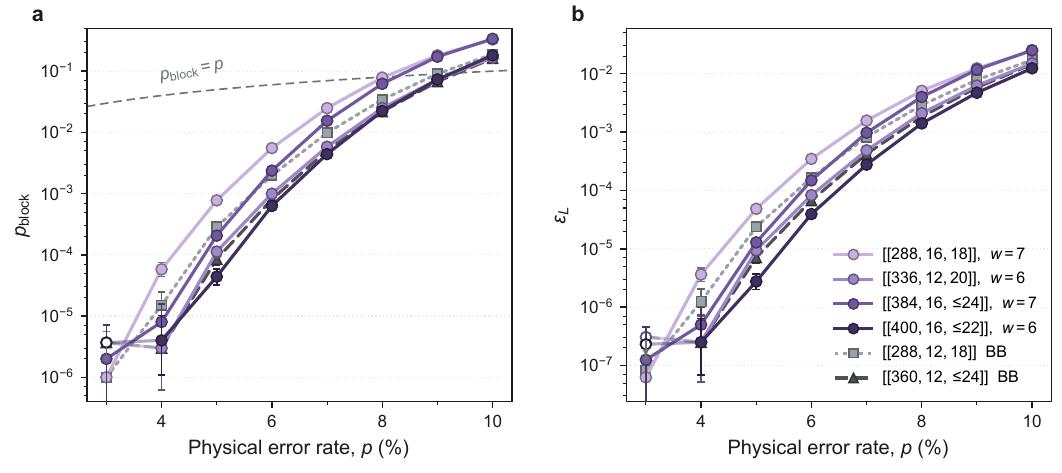}
  \caption{\textbf{Code-capacity performance under depolarizing noise.} (a)~Block error rate $p_{\mathrm{block}}$ and (b)~per-logical error rate $\epsilon_L$ versus the physical error rate $p$ for four discovered codes at $w=6$ and $7$ (solid purple: \qcode{288}{16}{18}, \qcode{336}{12}{20}, \qcode{384}{16}{\leq24} and \qcode{400}{16}{\leq22}) and two weight-six bivariate-bicycle benchmarks (dashed gray: \qcode{288}{12}{18} and \qcode{360}{12}{\leq24}), all decoded with the same minimum-sum BP--OSD protocol. The dashed line in (a) marks $p_{\mathrm{block}}=p$, whose crossing defines the pseudo-threshold. Error bars are exact 95\% confidence intervals; points with no observed failures are plotted as open markers at their upper confidence limits.}
  \label{fig:code-capacity}
\end{figure}

We next evaluated the discovered codes under a common code-capacity depolarizing-noise model. All discovered and reference codes were decoded using the same minimum-sum BP--OSD implementation, with no code-specific tuning. We denote the block error rate by $p_{\mathrm{block}}$, defined as the probability that a decoding trial produces at least one logical error. To compare codes encoding different numbers of logical qubits, we further consider the per-logical error rate $\epsilon_L=1-(1-p_{\mathrm{block}})^{1/k}$. The computation details can be found in Methods. Figure~\ref{fig:code-capacity} compares four parameter-leading discoveries at $w=6$ and 7 with two representative weight-six bivariate-bicycle benchmarks, \qcode{288}{12}{18} and \qcode{360}{12}{\leq24}, under the same protocol \citep{Bravyi2024Nature}. The discovered codes perform comparably across the sampled physical error rates. Moreover, the \qcode{400}{16}{\leq22} code performs better than both benchmarks across all physical error rates.

Table~\ref{tab:code-capacity} collects the leading exact-distance and upper-bound discoveries from every weight class, including the parameter champions not displayed in Fig.~\ref{fig:code-capacity}. Their estimated pseudo-thresholds $p^\star$, defined by the crossing of $p_{\mathrm{block}}(p)$ with $p$, are all above 5\% and range from 5.40\% to 9.30\%. At $p=0.03$, all ten codes have $\epsilon_L\leq1.68\times10^{-5}$, and eight are below $10^{-6}$. These low per-logical error rates provide a complementary assessment of the discovered codes and further support their finite-length quality beyond the parameter score.

The table also reveals a systematic weight dependence. Although the larger weight classes contain candidates with substantially higher $Q$, the pseudo-threshold decreases monotonically. This trend is not specific to the discovered codes: under the same protocol, we verified that weight-ten generalized-bicycle code \qcode{254}{28}{14\text{--}20} has a pseudo-threshold of 5.44\% \citep{PanteleevKalachev2021Finite}. These results suggest that the present BP--OSD implementation is better matched to lower-weight Tanner graphs. Decoder algorithms adapted to denser or more structured graphs may therefore yield stronger performance for higher-weight codes.

\begin{table}[H]
\centering
\begin{threeparttable}
\caption{\textbf{Code-capacity performance of the parameter champions.} One exact-distance code and one upper-bound code are reported for each overall-weight class. Values marked with an asterisk correspond to zero observed block failures in $10^6$ shots; the reported values are the upper endpoints of the two-sided exact 95\% Clopper--Pearson confidence intervals transformed to $\epsilon_L$.}
\label{tab:code-capacity}
\small
\setlength{\tabcolsep}{4.2pt}
\begin{tabular}{@{}c l r r r r r@{}}
\toprule
$w$ & Code & $Q$ & $p^\star$ & $\epsilon_L(0.03)$ & $\epsilon_L(0.05)$ & $\epsilon_L(0.07)$ \\
\midrule
6 & \qcode{336}{12}{20} & 14.29 & 9.30\% & $(3.07\times10^{-7})^{*}$ & $9.42\times10^{-6}$ & $4.87\times10^{-4}$ \\
6 & \qcode{400}{16}{\leq22} & $\leq19.36$ & 9.26\% & $(2.31\times10^{-7})^{*}$ & $2.75\times10^{-6}$ & $2.77\times10^{-4}$ \\
7 & \qcode{288}{16}{18} & 18.00 & 8.03\% & $6.25\times10^{-8}$ & $4.85\times10^{-5}$ & $1.57\times10^{-3}$ \\
7 & \qcode{384}{16}{\leq24} & $\leq24.00$ & 8.28\% & $1.25\times10^{-7}$ & $1.29\times10^{-5}$ & $9.74\times10^{-4}$ \\
8 & \qcode{378}{32}{19} & 30.56 & 7.08\% & $1.25\times10^{-7}$ & $5.41\times10^{-5}$ & $2.10\times10^{-3}$ \\
8 & \qcode{336}{24}{\leq24} & $\leq41.14$ & 7.11\% & $1.67\times10^{-7}$ & $8.02\times10^{-5}$ & $2.72\times10^{-3}$ \\
9 & \qcode{288}{18}{18} & 20.25 & 6.28\% & $2.72\times10^{-6}$ & $5.26\times10^{-4}$ & $8.37\times10^{-3}$ \\
9 & \qcode{384}{18}{\leq28} & $\leq36.75$ & 6.55\% & $4.44\times10^{-7}$ & $2.35\times10^{-4}$ & $6.99\times10^{-3}$ \\
10 & \qcode{234}{28}{18} & 38.77 & 5.40\% & $1.68\times10^{-5}$ & $1.21\times10^{-3}$ & $1.01\times10^{-2}$ \\
10 & \qcode{390}{32}{\leq32} & $\leq84.02$ & 5.78\% & $8.13\times10^{-7}$ & $4.89\times10^{-4}$ & $8.76\times10^{-3}$ \\
\bottomrule
\end{tabular}
\end{threeparttable}
\end{table}
\section{Discussion}

In summary, our work makes both practical and methodological contributions. On the practical side, it identifies qLDPC codes with strong finite-length parameters under practical length and weight constraints, together with structurally distinct constructions. These results provide relevant candidates for experimental QEC while broadening the range of finite-length constructions available for further investigation. On the methodological side, our results show that an agentic framework can support both effective optimization and broad exploration in qLDPC code discovery, offering a concrete example of how AI systems can contribute to scientific discovery.

Our results also suggest several directions for future work. First, the score $Q_{\text{proxy}}$ summarizes the finite-length rate--distance trade-off, whereas operational performance additionally depends on the parity-check representation, decoder and noise model. In the present study, logical failure probabilities are evaluated for selected codes after the parameter-based search; future implementations could incorporate decoder feedback directly into the search loop \citep{Liu2026SCE}. The search could also include structural constraints beyond overall weight. For example, requiring Tanner graphs with girth of at least six would suppress the shortest cycles and may improve performance under BP-based decoders.

Second, the framework could be extended to other qLDPC construction languages. The current implementation focuses on coset-orbit balanced products, but the same search architecture can accommodate other families once a deterministic constructor and a compact program representation are provided. Recent constructions based on multivariate multicycle codes, pair partitions, systematic lifts of lifted-product codes and hardware-oriented routing codes provide natural directions for such extensions \citep{Mian2026Multicycle,OkadaKasai2026PairPartition,HirasakiLee2026Lifting,Zhang2026Routing}.

Finally, the framework could be applied beyond code discovery to other important tasks in quantum error correction and fault-tolerant quantum computation. More generally, tasks with a well-defined objective and candidate solutions that can be expressed as executable programs and assessed through computation or simulation are natural targets for this approach. Examples include discovering decoder algorithms \citep{Moradi2026ClusterBP}, developing faster methods for distance estimation and certification \citep{Webster2026Distance}, and designing fault-tolerant gadgets and hardware layouts \citep{HirasakiLee2026Lifting,He2026TeXRA,Mathews2026Hardware}. These extensions could support the joint design of codes, decoders and fault-tolerant implementations.
\section{Methods}

The subsections below follow the order in which the components of the framework are introduced in the main text. We first describe the researcher council, curator and worker, and then give the deterministic evaluation procedure applied to every worker submission. We next define the executable family representation and balanced-product construction used in the present search, followed by the MAP-Elites archive, task scheduler and feedback loop. The final subsections distinguish agent-directed distance verification during the search from post-search MILP certification, and then describe the published-code comparison and code-capacity simulations used in the results.

\subsection{Researcher council, curator and worker execution}
In the reported search, all agents were implemented using Claude Code; the researcher-council members and the curator used Opus~4.8, whereas the worker used Sonnet~5.

The researcher council generates theory-driven proposals for the search. Each of its three members receives a compact summary of the evaluated population, current lessons and representative candidates, together with access to relevant literature and read-only structural measurements. The researchers reason independently from the perspectives of finite groups, qLDPC codes, and distance and sparse-graph structure. They then cross-review one another's proposals and return revised proposals and distance-verification recommendations to the curator.

The curator combines the reviewed proposals with evidence generated by the search, including recent outcomes, informative parent--child and cross-family contrasts, worker notes and unresolved validation results. It records each lesson as a falsifiable statement linked to explicit supporting and contradicting candidates. To avoid counting related descendants as independent discoveries, candidates that share any search root form one evidence group. For a lesson with prior strength $p$, $s$ independent supporting groups and $c$ independent contradicting groups, the confidence is
\begin{equation}
C_{\mathrm{lesson}}=\frac{1+p+s}{2+p+s+c}.
\end{equation}
Positive lessons are labeled hypothesis, leaning or confirmed at confidence thresholds $0.60$ and $0.85$, while anti-lessons form a separate negative category. The status is recomputed from the cited candidates rather than supplied by an agent, allowing later counterexamples to demote an earlier conclusion. The curator synthesizes at most five reviewed directions into worker tasks and can also issue high-budget distance-verification requests as described below.

The worker executes one task at a time. It receives zero or more selected parent programs, the intended mutation level, active lessons and anti-lessons, relevant failed siblings, cross-niche exemplars and a compact lineage history. Read-only tools allow the worker to assemble a draft, test validity, check for duplicates and probe the search score. During a multi-round session, it can construct, test and revise several drafts, but only one final family program is submitted to the evaluator. Alongside that submission, the worker produces a short free-form note recording the variants tried, their probe or self-check results, the reason for its final choice, and any dead ends or promising leads. This note is stored with the candidate and passed to later workers in the lineage and to the curator. Worker sessions in the reported search were limited to 40 agent turns.

\subsection{Deterministic evaluation and family score}
The deterministic evaluator applies the same sequence of checks to every submitted family program. It scans the program's declared values of $t$ within the global range $1\leq t\leq500$, skips groups of order above 20{,}000 and retains only assembled matrices in the window $40\leq n\leq400$. All code parameters are recomputed from the binary matrices. A parameter point is valid only if it satisfies $H_X H_Z^{\mathsf T}=0$, encodes
\begin{equation}
k=n-\operatorname{rank}(H_X)-\operatorname{rank}(H_Z)>0
\end{equation}
logical qubits and obeys the overall-weight constraint in Eq.~\eqref{eq:weight}.

From the valid parameter points, the evaluator retains at most three representatives for full scoring. It first selects the point with the largest $k$, then the highest-rate point among those remaining, and finally the next-largest k among the remaining points; repeated selections are suppressed. Each retained matrix is converted to a three-colored Tanner graph with separate colors for qubits, $X$ checks and $Z$ checks. BLISS canonical labeling gives an exact isomorphism key under color-preserving permutations~\cite{junttila2007engineering}. A representative is marked as a duplicate and assigned zero score if an isomorphic matrix is already present in the archive.

Every new, non-certified representative first receives a QDistEvol probe with \texttt{iterCount} set to $10^4$ and \texttt{genCount} to 1{,}000~\cite{Webster2026Distance}. If the resulting logical-operator witness gives $d_{\mathrm{ub}}>1.3\sqrt n$, the evaluator automatically repeats the calculation with \texttt{iterCount} set to $10^5$ and \texttt{genCount} to 2{,}000. The point score is then computed using Eq.~\eqref{eq:proxy-score}; its cap prevents a loose upper bound from being rewarded as if it were a genuine distance improvement~\cite{CruzBenito2026Evolution}. A previously certified exact distance can be used directly instead of the capped upper bound. The family score is the largest point score among the fully evaluated representatives, and the representative attaining this maximum supplies the rate and weight coordinates used by the archive. The separate $10^6$-iteration verification tier is described below.

\subsection{Executable family programs and balanced-product construction}
The unit of evolution is a program that generates an entire code family rather than a single parity-check matrix. For each admissible value of the scaling parameter $t$, the program constructs a finite group $G(t)$ and specifies two protographs $A(t)$ and $B(t)$. It defines their shapes and scaling with $t$, the subgroups assigned to their row and column types, and the local terms in every entry. The subgroup assignment may use one common subgroup $K$ or distinct subgroups for different row and column types.

Each local term is an $\mathbb F_2$ sum of double-coset orbits
\begin{equation}
K_i g K_j=\{k_i g k_j:\,k_i\in K_i,\ k_j\in K_j\},
\end{equation}
where $K_i$ and $K_j$ are associated with the corresponding protograph types. Before assembly, every entry is projected onto canonical double-coset representatives, and repeated orbits cancel modulo two. The deterministic assembler then materializes the required coset spaces and applies the balanced-product incidence rules to obtain $H_X(t)$ and $H_Z(t)$.

\subsection{MAP-Elites archive, scheduling and feedback}
The MAP-Elites archive stores evaluated families and supplies parents for subsequent tasks. Its niches are indexed by host-group family, subgroup structure, rate bin and overall-weight class~\cite{mouret2015illuminating}. The subgroup coordinate distinguishes trivial, normal, non-normal and mixed assignments. Writing $r=k/n$ for the rate of the best-scoring representative, the five rate bins have lower edges $0$, $0.05$, $0.10$, $0.15$ and $0.20$. A family replaces the current elite of its niche only when its family score is larger.

The scheduler first considers tasks proposed by the researcher council and curator in descending order of priority. The first task whose parent conditions can be met is paired with the highest-scoring matching parent or parents; a task explicitly marked as parentless bypasses this matching step. Tasks that test an existing lesson are subject to a $10\%$ dispatch-budget guard. If no queued task can be dispatched, the scheduler creates an automatic refinement task from the archive.

Automatic parent selection proceeds from a weight class to a niche and then to an individual family. Let $\mathcal C$ denote the non-empty classes among $6,\ldots,10$, and let $M_w$ be the number of selectable niches in class $w$. The weight class is sampled according to
\begin{equation}\label{eq:weight-sampling}
P(w)=\frac{\epsilon}{|\mathcal C|}
 +(1-\epsilon)\frac{\sqrt{1+M_w}}
 {\sum_{v\in\mathcal C}\sqrt{1+M_v}},
\qquad \epsilon=0.3.
\end{equation}
The first term guarantees exposure for every available weight class, while the second allocates more trials to classes containing more distinct niches, but only sublinearly in their number.

Within the selected weight class, niche selection balances quality, coverage and freshness. The unnormalized weight of niche $j$ is
\begin{equation}\label{eq:niche-sampling}
W_j=q_j c_j f_j,
\qquad
q_j=\exp\!\left[\frac{\phi(F_j)-\max_\ell\phi(F_\ell)}{\tau}\right],
\qquad
c_j=1+\beta s_j ,
\end{equation}
where $F_j$ is the family score and
\begin{equation}
\phi(F)=\frac{\log(1+F)}{\log(1+F_{\max})}.
\end{equation}
Here $F_{\max}$ is the largest family score observed so far, with a minimum normalization scale of one. Thus $q_j$ is an annealed quality factor. The coverage factor $c_j$ uses $\beta=0.5$ and
\begin{equation}
s_j=\frac{1}{2}\left[
  \left(1-\frac{a_j}{a_{\max}}\right)
 +\left(1-\frac{b_j}{b_{\max}}\right)\right],
\end{equation}
where $a_j$ and $b_j$ count occupied niches sharing the group-family and rate-bin coordinates of niche $j$, respectively. $a_{\max}$ and $b_{\max}$ are the corresponding maxima within the weight class. Sparsely represented group families and rate bins therefore receive a larger coverage factor.

Freshness prevents a small set of parents from repeatedly consuming the search budget. For candidate $i$, it is
\begin{equation}
f_i=\frac{1}{\sqrt{1+D_i+B_i}},
\end{equation}
where $D_i$ is the previous dispatch count and $B_i$ is an unproductive-history penalty. The niche-level factor $f_j$ is the largest freshness among its eligible members. After a niche is selected, its eligible members are sampled using the same annealed quality factor multiplied by $f_i$.

The history penalty records whether previous descendants were informative. A child that exceeds its parent or becomes an elite resets $B_i$ to zero without resetting $D_i$. An invalid or zero-score child adds two units, whereas a valid child that improves neither its parent nor any niche adds one. If an elite reaches $B_i\geq4$, the strongest runner-up with $B_i<4$ becomes eligible when its score is at least $0.3$ of the elite score.

Finally, the scheduler samples a mutation level for an automatic refinement task. The temperature in Eq.~\eqref{eq:niche-sampling} decreases geometrically from $\tau=0.6$ to $0.3$ over the course of the campaign. Over the same interval, the L1--L4 probabilities interpolate geometrically from $(1/3,1/3,1/6,1/6)$ to $(0.50,0.30,0.08,0.12)$ and are normalized before sampling. The schedule gradually shifts trials from structural L3/L4 exploration toward L1/L2 refinement while retaining a non-zero probability for every level.

After evaluation, the child updates the appropriate archive niche and is compared with its parent or parents. The resulting parent--child and cross-family contrasts, together with worker notes and validation outcomes, become evidence for the curator and researcher council. In the reported search, the curator was invoked after every 25 evaluated candidates. The researcher council normally convened after every 50 candidates and was also triggered after eight consecutive unsuccessful evaluations. Their new tasks and lessons then re-enter the scheduler, closing the search loop.

\subsection{Agent-directed distance verification}
The $10^6$-iteration verification tier is distinct from the automatic $10^5$-iteration escalation used for scoring. It is not triggered by a numerical threshold. Instead, the curator can issue a dedicated verification request when a tighter distance bound may resolve a lesson, distinguish competing explanations or change a selection decision. A request can arise from the curator's own contrastive analysis or from a recommendation made by the researcher council. At most three requests are admitted in one curator cycle.

QDistEvol, an evolutionary descendant of QDistRnd~\cite{pryadko2022qdistrnd}, is run for each request with \texttt{iterCount} set to $10^6$ and \texttt{genCount} to $10^4$. The $X$ and $Z$ sectors are searched independently with seeds 1001--1004, giving eight runs per code. Every run returns an explicit logical-operator witness, which is checked by independent binary linear algebra. The tightest validated upper bound is written back to the candidate, after which its family score and archive position are recomputed and the new evidence becomes available to the curator. These calculations can tighten an upper bound but cannot certify the distance.

\subsection{Post-search MILP distance certification}
MILP certification is applied after the search to selected code matrices. For a code with $k$ logical qubits, we choose a symplectic basis of logical generators and solve one mixed-integer minimization for each generator in each Pauli sector. A subproblem minimizes the Hamming weight of an operator that commutes with the relevant checks and anticommutes with the chosen dual logical generator. Any nontrivial logical operator anticommutes with at least one dual generator, so these $2k$ overlapping subproblems cover the full logical group. QDistEvol witnesses warm-start the solver and provide validated upper bounds for the minimizations.

A distance is certified as exact only when an explicit witness supplies the global upper bound and every subproblem has terminated with a proven optimum or has proved that no operator up to that bound is feasible. Otherwise, it is not reported as exact. Evidence bundles store matrix hashes, every solver status and the optimal witnesses, which are replayed using independent $\mathbb F_2$ arithmetic. Exact claims in this paper therefore rely on solver-attested global optimality under Gurobi version 12.0.3~\cite{gurobi}.

\subsection{Comparison with published codes}
The gray points in Fig.~\ref{fig:parameter-landscape} come from a curated comparison set of published binary CSS and stabilizer LDPC instances with $n\leq400$ and $k>0$. The literature benchmark comprises 1,209 code instances from 91 sources identified before the completion of this work.

Each entry is assigned one of five distance-evidence labels: exact distance, certified lower bound, upper bound, two-sided interval or unverified numerical claim. Exact values and upper bounds are compared separately. The overall weight is recomputed from the defining matrices or construction data using Eq.~\eqref{eq:weight}. 

\subsection{Code-capacity simulations}
The code-capacity protocol applies independent depolarizing noise to data qubits, with probabilities $1-p,p/3,p/3,p/3$ for $I,X,Y,Z$. The $X$ and $Z$ sectors are decoded independently using minimum-sum belief propagation followed by order-four combination-sweep ordered-statistics decoding \citep{PanteleevKalachev2021Finite,roffe2020decoding}. Belief propagation uses a parallel schedule, 5,000 iterations and scaling factor 1.0. The implementation uses the \texttt{ldpc} package \citep{roffe2022ldpc}. Physical error probabilities are $p=0.02,0.03,\ldots,0.10$. All codes use the same implementation, channel prior, seed policy and decoder settings; no parameters are tuned for an individual code.

Each point uses at most $10^6$ shots. Simulations of non-BB codes stop early after exactly 10,000 block errors, whereas the bivariate-bicycle references use $10^6$ shots throughout. We report the block error rate $p_{\mathrm{block}}$, defined as the probability that a decoding trial produces at least one logical error, and the per-logical error rate
\begin{equation}
\epsilon_L=1-(1-p_{\mathrm{block}})^{1/k}.
\end{equation}
Fixed-shot points use exact Clopper--Pearson 95\% confidence intervals. Adaptively stopped points use the corresponding exact negative-binomial intervals, and zero-failure observations are displayed as upper limits. The pseudo-threshold $p^\star$ solves $p_{\mathrm{block}}(p^\star)=p^\star$ and is estimated by log-linear interpolation between the two adjacent simulated points that bracket the crossing.

\section*{Data availability}

Explicit constructions of all discovered codes presented in the main text are given in the Supplementary Information. All other data, including detailed recoreds for all code instances for comparison, will be made publicly available upon publication.

\section*{Code availability}

The construction, search and evaluation code will be made publicly available upon publication.

\bibliographystyle{unsrt}

\clearpage
\section*{Supplementary Information}
\subsection*{Construction data for the codes in Tables~\ref{tab:codes} and \ref{tab:structural-codes}}
Each code below is the coset-orbit balanced product specified by its host group $G$, stabilizer subgroup $K$ and protograph matrices $A$ and $B$, assembled as described in Methods. Every entry of $A$ and $B$ is an $\mathbb F_2$ sum of double cosets $KgK$, written through the listed representatives $g$; for $K=\{e\}$ the entries are ordinary group-algebra elements. Group elements are written multiplicatively in the stated generators, with $e$ the identity.

\paragraph{$\qcode{336}{12}{20}$, $w=6$ (Table~\ref{tab:codes}).}
$G=\mathbb Z_{84}\rtimes_{29}\mathbb Z_{4}$, generators $x,s$: $x^{84}=s^{4}=e$, $sxs^{-1}=x^{29}$; $|G|=336$. $K=\langle x^{42}\rangle$ (normal, order 2).
\begin{align*}
A &= e+xs+x^{3}, &
B &= s^{3}+x^{2}+x^{4}.
\end{align*}

\paragraph{$\qcode{400}{16}{\leq22}$, $w=6$ (Table~\ref{tab:codes}).}
$G=\mathrm{Dic}_5\times\mathrm{Dic}_5$, generators $r_i,s_i$ ($i=1,2$): $r_i^{10}=e$, $s_i^{2}=r_i^{5}$, $s_ir_is_i^{-1}=r_i^{-1}$; $|G|=400$. $K=\langle r_1^{5}r_2^{5}\rangle$ (normal, order 2).
\begin{align*}
A &= s_1r_2s_2+r_1^{9}r_2^{2}s_2+r_1^{2}s_1r_2^{3}s_2, &
B &= r_1^{8}+r_1^{2}r_2+r_2s_2.
\end{align*}

\paragraph{$\qcode{288}{16}{18}$, $w=7$ (Table~\ref{tab:codes}).}
$G=\mathbb Z_{12}\times\mathbb Z_{48}$, generators $x,y$: $x^{12}=y^{48}=e$; $|G|=576$. $K=\langle y^{12}\rangle$ (normal, order 4).
\begin{align*}
A &= y^{2}+y^{7}+x, &
B &= y^{3}+x+x^{2}+x^{5}y^{9}.
\end{align*}

\paragraph{$\qcode{384}{16}{\leq24}$, $w=7$ (Table~\ref{tab:codes}).}
$G=\mathbb Z_{12}\times\mathbb Z_{48}$, generators $x,y$: $x^{12}=y^{48}=e$; $|G|=576$. $K=\langle x^{4}\rangle$ (normal, order 3).
\begin{align*}
A &= y^{14}+y^{19}+x, &
B &= y^{3}+x+x^{2}+x^{5}y^{21}.
\end{align*}

\paragraph{$\qcode{378}{32}{19}$, $w=8$ (Table~\ref{tab:codes}).}
$G=\mathrm{He}(\mathbb Z_3)\times\mathbb Z_{7}$, generators $x,y,z,u$: $x^{3}=y^{3}=z^{3}=u^{7}=e$, $z=xyx^{-1}y^{-1}$ central in $\mathrm{He}(\mathbb Z_3)$, $u$ generates $\mathbb Z_7$; $|G|=189$. $K=\{e\}$ (free action).
\begin{align*}
A &= zu^{3}+y+xu+xyu^{6}, &
B &= zu^{4}+yu^{5}+x+xyz^{2}u^{3}.
\end{align*}

\paragraph{$\qcode{288}{24}{18}$, $w=8$ (Table~\ref{tab:codes}).}
$G=\mathbb Z_{12}\times\mathbb Z_{48}$, generators $x,y$: $x^{12}=y^{48}=e$; $|G|=576$. $K=\langle y^{12}\rangle$ (normal, order 4).
\begin{align*}
A &= y^{2}+y^{7}+xy+x^{3}+x^{11}y^{11}, &
B &= y^{3}+x+x^{2}.
\end{align*}

\paragraph{$\qcode{224}{22}{16}$, $w=8$ (Table~\ref{tab:codes}).}
$G=\mathbb Z_{84}\rtimes_{29}\mathbb Z_{4}$, generators $x,s$: $x^{84}=s^{4}=e$, $sxs^{-1}=x^{29}$; $|G|=336$. $K=\langle x^{28}\rangle$ (normal, order 3).
\begin{align*}
A &= s^{2}+x+x^{2}s^{3}+x^{5}s^{2}, &
B &= s+x^{5}+x^{17}s+x^{20}s^{3}.
\end{align*}

\paragraph{$\qcode{336}{24}{\leq24}$, $w=8$ (Table~\ref{tab:codes}).}
$G=\mathbb Z_{4}\times\mathbb Z_{42}$, generators $x,y$; $|G|=168$. $K=\{e\}$ (free action).
\begin{align*}
A &= e+xy^{2}+x^{2}y^{5}+x^{3}y^{6}, &
B &= e+xy^{13}+x^{2}y^{9}+x^{3}y^{19}.
\end{align*}

\paragraph{$\qcode{378}{18}{\leq27}$, $w=8$ (Table~\ref{tab:codes}).}
$G=M_{27}\times\mathbb Z_{7}$, generators $x,s,u$: $x^{9}=s^{3}=u^{7}=e$, $sxs^{-1}=x^{4}$, $u$ generates $\mathbb Z_7$; $|G|=189$. $K=\{e\}$ (free action).
\begin{align*}
A &= e+s^{2}u^{3}+x^{3}su+x^{5}u^{6}, &
B &= s+x+x^{2}s^{2}u^{5}+x^{5}su^{3}.
\end{align*}

\paragraph{$\qcode{336}{28}{\leq20}$, $w=8$ (Table~\ref{tab:codes}).}
$G=\mathbb Z_{2}\times\mathbb Z_{2}\times\mathbb Z_{42}$, generators $x,y,z$; $|G|=168$. $K=\{e\}$ (free action).
\begin{align*}
A &= e+yz^{5}+xz^{2}+xyz^{6}, &
B &= e+yz^{9}+xz^{13}+xyz^{19}.
\end{align*}

\paragraph{$\qcode{288}{18}{18}$, $w=9$ (Table~\ref{tab:codes}).}
$G=\mathrm{SL}(2,3)\rtimes_{\theta}\mathbb Z_{12}$: elements $My^{z}$ with $M\in\mathrm{SL}(2,3)$, $y$ generating $\mathbb Z_{12}$ and $yMy^{-1}=\theta(M)=PMP^{-1}$, $P=\left(\begin{smallmatrix}1&0\\0&2\end{smallmatrix}\right)$; $|G|=288$. $K=\langle -I\rangle$ (normal, order 2).
\begin{align*}
A &= \left(\begin{smallmatrix}1&1\\0&1\end{smallmatrix}\right)+y+\left(\begin{smallmatrix}0&1\\2&0\end{smallmatrix}\right)y^{2}+\left(\begin{smallmatrix}1&0\\1&1\end{smallmatrix}\right)y^{3}, &
B &= \left(\begin{smallmatrix}1&2\\0&1\end{smallmatrix}\right)+\left(\begin{smallmatrix}2&1\\2&0\end{smallmatrix}\right)y+y^{2}+\left(\begin{smallmatrix}1&0\\2&1\end{smallmatrix}\right)y^{3}+\left(\begin{smallmatrix}2&1\\2&0\end{smallmatrix}\right)y^{6}.
\end{align*}

\paragraph{$\qcode{320}{24}{16}$, $w=9$ (Table~\ref{tab:codes}).}
$G=\mathbb Z_{41}\rtimes\mathbb Z_{8}$, generators $x,s$: $x^{41}=s^{8}=e$, $sxs^{-1}=x^{38}$; $|G|=328$. $K=\langle x\rangle=\mathbb Z_{41}$ (normal).
\begin{align*}
A &= \begin{pmatrix}e & x & s & s^{6}\\ s^{6} & e & x & s\\ s & s^{6} & e & x\\ x & s & s^{6} & e\end{pmatrix}, &
B &= \begin{pmatrix}e & x^{5}s^{3} & s^{2} & x^{15}s^{3} & x^{20}s^{3}\\ x^{20}s^{3} & e & x^{5}s^{3} & s^{2} & x^{15}s^{3}\\ x^{15}s^{3} & x^{20}s^{3} & e & x^{5}s^{3} & s^{2}\\ s^{2} & x^{15}s^{3} & x^{20}s^{3} & e & x^{5}s^{3}\\ x^{5}s^{3} & s^{2} & x^{15}s^{3} & x^{20}s^{3} & e\end{pmatrix}.
\end{align*}

\paragraph{$\qcode{400}{26}{16}$, $w=9$ (Table~\ref{tab:codes}).}
$G=\mathrm{Dic}_4$, generators $r,s$: $r^{8}=e$, $s^{2}=r^{4}$, $srs^{-1}=r^{-1}$; $|G|=16$. $K=\{e\}$ (free action).
\begin{align*}
A &= \begin{pmatrix}e & r & r^{2}s & r^{5}\\ r^{2}s & e+r^{4} & r & r^{4}s\\ r & r^{7}s & e & r^{2}\end{pmatrix}, &
B &= \begin{pmatrix}e & r^{2} & r^{5}s & r\\ rs & e & r^{2} & r^{4}\\ r^{3} & r^{5}s & e & r^{5}\end{pmatrix}.
\end{align*}

\paragraph{$\qcode{384}{18}{\leq28}$, $w=9$ (Table~\ref{tab:codes}).}
$G=\mathrm{SL}(2,3)\rtimes_{\phi}\mathbb Z_{16}$: elements $My^{z}$ with $M\in\mathrm{SL}(2,3)$, $y$ generating $\mathbb Z_{16}$ and $yMy^{-1}=\phi(M)$, where $\phi$ is conjugation by $\left(\begin{smallmatrix}1&1\\2&1\end{smallmatrix}\right)\in\mathrm{GL}(2,3)$; $|G|=384$. $K=\langle -I\rangle$ (normal, order 2).
\begin{align*}
A &= e+\left(\begin{smallmatrix}1&1\\1&2\end{smallmatrix}\right)y+\left(\begin{smallmatrix}1&0\\1&1\end{smallmatrix}\right)y^{3}+\left(\begin{smallmatrix}1&1\\0&1\end{smallmatrix}\right)y^{3}+\left(\begin{smallmatrix}0&1\\2&0\end{smallmatrix}\right)y^{6}, &
B &= e+\left(\begin{smallmatrix}1&2\\0&1\end{smallmatrix}\right)y+\left(\begin{smallmatrix}1&0\\1&1\end{smallmatrix}\right)y^{2}+\left(\begin{smallmatrix}0&1\\2&0\end{smallmatrix}\right)y^{5}.
\end{align*}

\paragraph{$\qcode{384}{14}{\leq28}$, $w=9$ (Table~\ref{tab:codes}).}
$G=\mathrm{GL}(2,3)\times\mathbb Z_{4}$: elements $My^{z}$ with $M\in\mathrm{GL}(2,3)$ and $y$ generating $\mathbb Z_{4}$; $|G|=192$. $K=\{e\}$ (free action).
\begin{align*}
A &= \left(\begin{smallmatrix}0&1\\1&0\end{smallmatrix}\right)+e+\left(\begin{smallmatrix}0&1\\2&0\end{smallmatrix}\right)y+\left(\begin{smallmatrix}1&1\\0&1\end{smallmatrix}\right)y^{2}+\left(\begin{smallmatrix}2&0\\0&2\end{smallmatrix}\right)y^{3}, &
B &= e+\left(\begin{smallmatrix}1&0\\1&1\end{smallmatrix}\right)y+\left(\begin{smallmatrix}1&1\\0&1\end{smallmatrix}\right)y^{2}+\left(\begin{smallmatrix}0&1\\2&0\end{smallmatrix}\right)y^{3}.
\end{align*}

\paragraph{$\qcode{234}{28}{18}$, $w=10$ (Table~\ref{tab:codes}).}
$G=\mathbb Z_{13}\times\mathbb Z_{9}$, generators $x,y$; $|G|=117$. $K=\{e\}$ (free action).
\begin{align*}
A &= e+y^{2}+y^{8}+x^{4}y^{4}+x^{6}y^{8}, &
B &= x^{2}y^{8}+x^{5}y^{4}+x^{10}y^{2}+x^{11}y^{5}+x^{12}.
\end{align*}

\paragraph{$\qcode{372}{44}{18}$, $w=10$ (Table~\ref{tab:codes}).}
$G=\mathbb Z_{31}\times\mathbb Z_{6}$, generators $x,y$; $|G|=186$. $K=\{e\}$ (free action).
\begin{align*}
A &= x^{5}y+x^{5}y^{3}+x^{7}y^{2}+x^{18}y^{2}+x^{30}y^{2}, &
B &= x^{10}y+x^{10}y^{3}+x^{21}y^{5}+x^{24}y^{5}+x^{26}y^{5}.
\end{align*}

\paragraph{$\qcode{170}{32}{14}$, $w=10$ (Table~\ref{tab:codes}).}
$G=\mathbb Z_{85}$, generator $x$; $|G|=85$. $K=\{e\}$ (free action).
\begin{align*}
A &= e+x^{3}+x^{4}+x^{10}+x^{67}, &
B &= e+x^{5}+x^{29}+x^{31}+x^{37}.
\end{align*}

\paragraph{$\qcode{390}{32}{\leq32}$, $w=10$ (Table~\ref{tab:codes}).}
$G=\mathbb Z_{195}$, generator $x$; $|G|=195$. $K=\{e\}$ (free action).
\begin{align*}
A &= e+x^{5}+x^{28}+x^{155}+x^{186}, &
B &= e+x+x^{64}+x^{86}+x^{161}.
\end{align*}

\paragraph{$\qcode{390}{36}{\leq30}$, $w=10$ (Table~\ref{tab:codes}).}
$G=\mathbb Z_{39}\times\mathbb Z_{5}$, generators $x,y$; $|G|=195$. $K=\{e\}$ (free action).
\begin{align*}
A &= e+x+x^{7}y+x^{8}y^{3}+x^{25}y^{2}, &
B &= e+xy^{2}+x^{3}y^{3}+x^{11}y+x^{12}y^{2}.
\end{align*}

\paragraph{$\qcode{256}{18}{16}$, $w=8$ (Table~\ref{tab:structural-codes}).}
$G=\mathbb Z_{8}\wr\mathbb Z_{2}$, generators $x_1,x_2,\sigma$: $x_1^{8}=x_2^{8}=\sigma^{2}=e$, $x_1x_2=x_2x_1$, $\sigma x_1\sigma=x_2$; $|G|=128$. $K=\{e\}$ (free action).
\begin{align*}
A &= x_2^{2}+x_1+x_1x_2^{6}\sigma+x_1^{2}x_2^{3}\sigma, &
B &= x_2\sigma+x_1x_2^{3}+x_1^{2}x_2^{4}\sigma+x_1^{5}.
\end{align*}

\paragraph{$\qcode{384}{32}{16}$, $w=8$ (Table~\ref{tab:structural-codes}).}
$G=\mathrm D_6\rtimes_{\theta}\mathbb Z_{8}$, generators $r,s,u$: $r^{6}=s^{2}=u^{8}=e$, $srs^{-1}=r^{-1}$, $uru^{-1}=r$, $usu^{-1}=r^{3}s$; $|G|=96$. $K=\langle r^{3}\rangle$ (normal, order 2).
\begin{align*}
A &= \begin{pmatrix}r+su^{2} & u+r^{4}u^{3}\\ u+r^{4}u^{3} & r+su^{2}\end{pmatrix}, &
B &= \begin{pmatrix}r+su^{3} & r^{2}u+r^{5}u^{2}\\ r^{2}u+r^{5}u^{2} & r+su^{3}\end{pmatrix}.
\end{align*}

\paragraph{$\qcode{368}{18}{16}$, $w=9$ (Table~\ref{tab:structural-codes}).}
$G=A_6\times\mathbb Z_{2}$: elements $\pi u^{z}$ with $\pi\in A_6$ (permutations of $\{1,\dots,6\}$, cycle notation) and $u$ generating $\mathbb Z_2$; $|G|=720$. $K=\langle(1\,2)(3\,4)\rangle$ (non-normal, order 2).
\begin{align*}
A &= (2\,3\,4\,5\,6)+(1\,2\,3)u, &
B &= (2\,3\,4\,5\,6)u+(2\,6\,5\,4\,3)u+(1\,3)(2\,4)u.
\end{align*}

\paragraph{$\qcode{336}{12}{\leq24}$, $w=9$ (Table~\ref{tab:structural-codes}).}
$G=\mathrm{PSL}(2,11)$: matrices over $\mathbb F_{11}$ with unit determinant, modulo $\pm I$ (one representative listed); $|G|=660$. $K=\langle\left(\begin{smallmatrix}0&1\\10&0\end{smallmatrix}\right)\rangle$ (non-normal, order 2).
\begin{align*}
A &= \left(\begin{smallmatrix}1&0\\4&1\end{smallmatrix}\right)+\left(\begin{smallmatrix}1&1\\0&1\end{smallmatrix}\right), &
B &= e+\left(\begin{smallmatrix}1&0\\4&1\end{smallmatrix}\right)+\left(\begin{smallmatrix}1&10\\0&1\end{smallmatrix}\right).
\end{align*}

\paragraph{$\qcode{248}{12}{18}$, $w=10$ (Table~\ref{tab:structural-codes}).}
$G=\mathrm{PSL}(2,13)$: matrices over $\mathbb F_{13}$ with unit determinant, modulo $\pm I$ (one representative listed); $|G|=1092$. $K=\langle\left(\begin{smallmatrix}3&0\\0&9\end{smallmatrix}\right)\rangle$ (non-normal, order 3).
\begin{align*}
A &= \left(\begin{smallmatrix}1&0\\4&1\end{smallmatrix}\right)+\left(\begin{smallmatrix}1&1\\0&1\end{smallmatrix}\right), &
B &= \left(\begin{smallmatrix}0&1\\12&0\end{smallmatrix}\right)+\left(\begin{smallmatrix}1&12\\0&1\end{smallmatrix}\right).
\end{align*}

\paragraph{$\qcode{396}{8}{\leq32}$, $w=10$ (Table~\ref{tab:structural-codes}).}
$G=\mathbb Z_{31}\rtimes\mathbb Z_{6}$, generators $x,s$: $x^{31}=s^{6}=e$, $sxs^{-1}=x^{26}$; $|G|=186$. $K=\langle s^{2}\rangle$ (non-normal, order 3).
\begin{align*}
A &= \begin{pmatrix}e & x & s\\ s & e & x\\ x & s & e\end{pmatrix}, &
B &= \begin{pmatrix}e & x^{3}s^{3} & s^{5}\\ s^{5} & e & x^{3}s^{3}\\ x^{3}s^{3} & s^{5} & e\end{pmatrix}.
\end{align*}

\paragraph{$\qcode{306}{8}{\leq25}$, $w=10$ (Table~\ref{tab:structural-codes}).}
$G=(\mathbb Z_{17}^{2})\rtimes_{\mathrm{swap}}\mathbb Z_{2}$, generators $x,y,\sigma$: $x^{17}=y^{17}=\sigma^{2}=e$, $xy=yx$, $\sigma x\sigma=y$; $|G|=578$. $K=\langle\sigma\rangle$ (non-normal, order 2).
\begin{align*}
A &= y^{4}+x+x^{3}y^{3}\sigma, &
B &= x^{2}y^{3}+x^{5}y+x^{6}y^{6}\sigma.
\end{align*}

\end{document}